\documentstyle [12pt,a4,epsf] {article}
\newcommand{\ie} {{\it i.e.}}
\newcommand {\s}{{\rm salt}}

\newcommand{\be} {\begin{equation}}
\newcommand{\ee} {\end{equation}}
\newcommand{\bea} {\begin{eqnarray}}
\newcommand{\eea} {\end{eqnarray}}

\begin{document}

\title {Theoretical Approaches to Neutral and Charged  Polymer Brushes}
\author {Ali Naji$^{1,2}$, Christian Seidel$^3$, Roland R. Netz$^{1,2,\#}$ \\
     $^1$   Dept. of Physics, Technical University of Munich\\
James Franck Str., 85478 Garching, Germany\\
$^2$  Dept. of Physics, Ludwig-Maximilians-University\\
         Theresienstr. 37, 80333 Munich, Germany\\
    $^3$ Max-Planck-Institute for Colloids and Interfaces\\
14424 Potsdam, Germany\\
   $^{\#}$    netz@ph.tum.de}
\date{June 2004}
\maketitle
\begin{abstract}
     
Neutral or charged  polymers that are densely end-grafted to surfaces form 
brush-like structures and are highly stretched 
under good-solvent conditions.
We discuss and compare relevant  results from  scaling 
models, self-consistent-field methods and MD simulation  techniques
and concentrate on the conceptual simple case of planar substrates.
For neutral polymers the main quantity of interest is the brush height
and the polymer density profile, which can be well predicted from 
self-consistent calculations and simulations.
Charged polymers (polyelectrolytes) are of practical importance since
they are soluble in water.  Counterion degrees of freedom determine the
brush behavior in a decisive way and lead to a strong and non-linear 
swelling of the brush.
\newline
{\bf Keywords}: Brushes, polyelectrolytes, scaling theory, 
self-consistent field theory, simulation techniques
\end{abstract}

\tableofcontents

\section*{Legend of Symbols}
$a$: Kuhn length or effective  monomer size \\
$d$: height of counterion layer\\
$f$: fractional charge of the chain $0<f<1$ \\
$ F$: free energy  in units of $k_BT$ (per chain or unit area) \\
$h$: height of brush\\
$k_B T$: thermal energy\\
$L$: contour length of a chain\\
$\ell_B=e^2/(4 \pi \varepsilon k_B T)$:  Bjerrum length\\
$N$: polymerization index\\
$R$: end-to-end polymer chain radius\\
$R_0$: end-to-end radius of an ideal polymer\\
$R_F$: Flory radius of a self-avoiding chain\\
$v_2$:  2nd virial coefficient of monomers in solution\\
$\kappa^{-1}$: Debye-H\"uckel screening length\\
$\nu$: Flory exponent for the polymer size\\
$\Pi$: osmotic pressure in units of $k_BT$\\
$\rho_a$: grafting density of a polymer  brush\\
$\rho(z)$: monomer density at distance $z$ from grafting  surface \\
$\sigma$: Lennard-Jones diameter in simulation

\section{Introduction}
Polymers are long, chain-like molecules that consist of repeating subunits, the so-called
monomers\cite{Yamakawa,degennes,Grosberg,Rubinstein}. 
In many situations,  all monomers of a polymer are
alike,  showing for example the same tendency to adsorb  to a
substrate\cite{Netzreview}. For industrial applications,
one is often interested  in {\em end-functionalized polymers} that 
are attached with one end  only to the
substrate\cite{halperin92,Szleifer}.
Industrial interest comes from
the need to stabilize particles  and surfaces against
flocculation. In end-grafted polymer structures the stabilization power
is greatly enhanced 
 as compared with adsorbed layers of polymers, where each
monomer is equally attracted to the substrate. The main reason is that 
bridging of polymers between two approaching surfaces
and creation of polymer loops on the same surface is very frequent 
in the case of polymer adsorption and eventually leads
to attraction between two particle surfaces and thus destabilization.
This does  not occur if the polymer is grafted by its end to the surface
and the monomers are  chosen such
that they  do not particularly adsorb to the surface.
In biology, brush-like polymer structures are encountered as the surface
coating of endothelial cells and regulate the adsorption and migration
of various particles and bio-molecules  from the blood stream to the vascular cells.

Experimentally, two basic ways of building a 
grafted polymer  layer can be distinguished:
In the first, the polymerization is started from the surface
with some suitably chosen surface-linked initiator. The advantage
of this {\em grafting-from} procedure is that only monomers have 
to diffuse through the forming brush layer and thus the 
reaction kinetics is fast. In the second route one attaches polymers 
with special end-groups that act as anchors on the surface.
This {\em grafting-to} procedure is subject to slow kinetics during the 
formation stage since whole polymers have to diffuse through
the natant grafting layer, but benefits from a somewhat
better control over the  brush constitution and chemical composition. 
One distinguishes {\em physical} adsorption of end-groups that favor the substrate,
for example zwitter-ionic end-groups attached to poly-styrene chains that 
lead to binding to mica in organic solvents such as toluene or xylene\cite{taunton}.
A stronger and thus more stable attachment is possible with {\em covalently} end-grafted chains, 
for example poly-dimethylsiloxane chains which carry hydroxyl end groups and
undergo condensation reactions with silanols of a silica surface\cite{auroy1}.
One can also employ a  suitably chosen diblock copolymer where
one block adsorbs on the substrate while the other is repelled from it\cite{marques}. 
An example is furnished by polystyrene-poly(vinyl-pyridine) (PS-PVP) diblocks
in the selective solvent toluene, which is a bad solvent for the PVP block
and promotes strong adsorption on a quartz substrate, but acts as a good solvent 
for the PS block and disfavors its adsorption on the substrate\cite{field}.
In a slight modification one uses
diblock copolymers that are anchored at the
liquid-air~\cite{kent,bijster} or at a liquid-liquid interface of two
immiscible liquids~\cite{teppner}. This scenario offers the
advantage that the grafting density can (for the case of strongly
anchored polymers) be varied by lateral compression (like a Langmuir mono-layer) 
and that the lateral surface pressure can be directly measured. 
The lateral pressure is an important thermodynamic quantity and allows 
detailed comparison with theoretical predictions. 
A well studied example is that of a diblock copolymer of
polystyrene -- polyethylene oxide (PS-PEO)\cite{bijster}. 
The PS block is shorter and
functions as an anchor at the air/water interface because it is
immiscible in water. The PEO block is miscible in water but because
of attractive interaction with the air/water interface it forms a
quasi-two dimensional layer at very low surface coverage. As the
surface pressure  increases and the area per polymer decreases, the PEO
block is expelled from the surface and forms a polymer brush.

In this theoretical chapter 
we simplify the discussion by assuming  that the
polymers are irreversibly grafted at one of their chain ends to the substrate, 
we only mention in passing a few papers on the kinetics of grafting. 
 The substrate is  assumed to be solid, planar and impenetrable to the 
 the polymer monomers, we briefly cite some results obtained for
 curved substrates.  
We limit the discussion to good solvent conditions and neglect
any attractive interactions between the polymer chains and the surface.
Charged polymers
are interesting from the application point of view,
since they allow for  water-based formulations of organic substances which
are advantageous for economical and ecological reasons. Recent years
have seen a tremendous research activity on charged polymers in
bulk\cite{Oosawa,Foerster,barrat1,holmadv} and at interfaces\cite{Netzreview,ruhe}. 
We therefore treat neutral brushes  as well as charged ones. 

The characteristic parameter for brush systems 
 is the anchoring or grafting density 
$\rho_a$, which is the inverse of the average area
available for each polymer at the surface. For small
grafting densities, $\rho_a < \rho_a^*$, the polymer chains will be far
apart from each other and hardly interact, as schematically shown
in Fig.1a. The polymers in this case form well separated 
{\em mushrooms} at the surface.
The grafting density at which chains just start to overlap is 
determined by $\rho_a^* \sim R^{-2}$ where $R$ is the typical
radius or size of a chain. 
In good solvent conditions (that is for swollen chains), the
chain radius follows  the Flory prediction 
$R_F  \sim a N^{3/5}$ where $N$ is the polymerization index or
monomer number of the chain, and $a$ is a characteristic 
microscopic length scale of the polymer that incorporates 
monomer size as well as backbone stiffness.
The crossover grafting density for a polymer under good-solvent
conditions follows as 
$\rho_a^*  \sim  a^{-2}N^{-6/5}$. 
For large grafting densities, $\rho > \rho^*$, the chains
are strongly  overlapping. 
 This situation is depicted in Fig.1b.
 Since we assume the solvent to be good, monomers repel
each other. The lateral separation between the polymer coils is
fixed by the grafting density, so that the polymers extend away
from the grafting surface in order to avoid each other.
The resulting structure is called a {\em polymer
brush}, with a vertical height $h$ which greatly exceeds the
unperturbed coil radius $R$~\cite{dolan,alex,gennes}.
Similar stretched
structures occur in many other situations, such as diblock
copolymer melts in the strong segregation regime~\cite{sem,halperin92}, 
or star polymers under good solvent conditions\cite{daoudcotton}. 
Theory is mostly concerned with predicting the layer height $h$, but
also the detailed monomer density profile and resulting forces
of lateral or vertical brush compression as a function of the various
system parameters.

The understanding of grafted 
polymer systems progressed  substantially with
the advent of   experimental techniques such as:
surface-force-balance\cite{taunton},
small-angle-neutron scattering\cite{auroy1},
neutron\cite{field,karim} and X-ray\cite{baltes} diffraction, and
ellipsometry\cite{teppner}.
Of equal merit was the advancement in the  theoretical
methodology 
ranging from field theoretical methods and scaling arguments
to numerical simulations, which will be amply reviewed in this chapter.

\section{Polymer basics}

The main parameters used to describe a polymer chain are
the polymerization index $N$, which counts the number of repeat
units or monomers along the chain, and 
the  size of one monomer or the distance between two neighboring
monomers. The  monomer size ranges from  a few Angstroms for synthetic
polymers to a few nanometers for biopolymers.
The simplest theoretical description
of flexible chain conformations is achieved
with the so-called freely-jointed chain (FJC) model,
where a polymer consisting
of $N+1$ monomers is represented by $N$ bonds defined by bond
vectors ${\bf r}_j$ with $j = 1, \ldots N$. Each bond vector has
a fixed length $|{\bf r}_j| = a$ corresponding to the Kuhn
length, but otherwise is allowed to rotate freely and independently
of its neighbors, as is
schematically shown in Fig.2 (top). This model of course
only gives a coarse-grained description of real polymer chains,
but we will later see that by a careful interpretation of the Kuhn length $a$
and the monomer number $N$, an accurate description
of the large-scale properties of real polymer chains is possible.
The main advantage is that due to the simplicity of the FJC model,
many interesting
observables (such as chain size or distribution functions) can
be calculated with relative ease. We demonstrate this by calculating
the mean end-to-end radius of such a FJC polymer.
Fixing one of the chain
ends at the origin, the position of the $(k+1)$-th monomer is
given by the vectorial sum
 \be
  {\bf R}_k = \sum_{j=1}^k {\bf r}_j ~.
 \ee
Because  two arbitrary bond vectors are uncorrelated in this
simple model, the thermal average over the scalar product of two
different bond vectors vanishes, $\langle {\bf r}_j\cdot {\bf r}_k
\rangle = 0 $ for $j \neq k$, while the mean squared bond vector
length is simply given by $\langle {\bf r}_j^2 \rangle = a^2 $.
It follows that the mean squared end-to-end radius $R_N^2$  is
proportional to the number of monomers,
 \be \label{idealscaling}
 \label{fjc} R_0^2 \equiv \langle {\bf R}_N^2 \rangle = N a^2 = La ,
 \ee
where the contour length of the chain is given by $L= Na$. 
$R_0$ denotes the mean end-to-end radius of an ideal chain,
and according to Eq.(\ref{idealscaling}),
it scales as $R_0 = a N^{1/2}$.  Experimentally one often knows,
via knowledge of the chemical structure and polymer mass,
the length $L$ and, via light-scattering,  the radius $R$ of a chain. 
Using the above scaling results, valid for a non-interacting chain, the Kuhn length
follows as $a= R_0^2/L$ and the effective monomer number as
$N=L/a$, which allows treatment of  real chains with a complicated
local conformational structure  within the 
FJC model. Note that the so-determined Kuhn length $a$ is usually
larger than the actual monomer-size, since it takes back-bone stiffness
effects into account. Likewise, the determined effective monomer number $N$
is typically smaller than the actual (chemical) number of monomers in the chain. 

In many theoretical calculations aimed at elucidating
large-scale properties, the simplification is carried even
a step further and a continuous model is used, as schematically
shown in Fig.2 (bottom). In such models the polymer backbone is
replaced by a continuous line and all microscopic details are neglected.
The chain is then only characterized by its length $L$ and radius $R$.

\subsection{Polymer swelling and collapse}

The models discussed so far describe ideal chains
and do not account for interactions between monomers which typically
consist of some short-ranged repulsion and long-ranged attraction. 
Including these interactions will give a different scaling behavior for
long polymer chains. The end-to-end radius, $R=\sqrt{\langle
R^2_N\rangle}$, can be written for  $N\gg 1$ as
\be R \simeq a N^{\nu}
\ee
which defines the so-called {\em swelling exponent} $\nu$. As we have seen,
for an ideal polymer chain (no interactions between monomers),
Eq.(\ref{idealscaling})
implies $\nu = 1/2$. This situation is realized  for experimental 
polymers at a certain temperature or solvent conditions
when  the attraction between monomers exactly  cancels the steric repulsion
(which is due to the fact that the monomers cannot penetrate each
other). This situation can be achieved in the
condition of  {\em theta solvents}.
In {\em good solvents}, on the other hand, the monomer-solvent
interaction is more attractive  than the monomer-monomer interaction, in other
words, the monomers try to avoid each other in solution. As a consequence,
single polymer chains in good solvents have swollen spatial
configurations dominated by the steric repulsion, characterized by
an exponent $\nu \simeq 3/5$, leading to the Flory radius $R_F \sim a N^{3/5}$. 
This spatial size of
a polymer coil is much smaller than the extended contour length
$L=aN$ but larger than the size of an ideal chain $R_0=aN^{1/2}$,
i.e. $R_0 \ll R_F \ll L $ in the limit $N \rightarrow \infty$. The
reason for this behavior is conformational entropy (which prevents
full stretching of the chain) combined with the
favorable interaction between monomers and solvent molecules in
good solvents (which leads to a more open structure
than an ideal chain).
In the opposite case of  {\em bad  or poor solvent}
conditions, the effective interaction between monomers is
attractive, leading to collapse of the chains and to their
precipitation from solution. In this case, the
polymer volume, like any space filling object embedded in
three-dimensional space, scales proportional to its weight as $R^3 \sim N$, 
yielding an exponent $\nu=1/3$.

The standard way  of taking into account interactions between
monomers is the Flory theory, which treats these interactions on
an approximate mean-field level\cite{Yamakawa,degennes,Grosberg}. Let us first
consider the case of repulsive interactions between monomers,
which can be described by a positive second-virial
coefficient $v_2$ and  corresponds to the  good-solvent
condition. For pure hard-core interactions and with no additional
attractions between monomers, the second virial coefficient
(which corresponds to the excluded volume) is of the order of the monomer
volume, i.e. $v_2 \sim a^3$. The repulsive interaction between monomers, which
tends to swell the chain, is counteracted and balanced by the
ideal chain elasticity, which is brought about by the entropy
loss associated with  stretching the chain.
The origin is that the number
of polymer configurations having an end-to-end radius of the
order of the unperturbed end-to-end radius is large. These
configurations are entropically favored over
configurations characterized by a large end-to-end radius, for
which the number of possible polymer conformations is drastically
reduced. The standard Flory theory for a flexible
chain of radius $R$ is based on writing the free energy $F$ of a chain
(in units of the thermal energy $k_BT$)
as a sum of two terms (omitting numerical prefactors)
\be 
F \simeq  \frac{ R^2}{ R_0^2 } + v_2 R^3 \left(
\frac{N }{R^3} \right)^2 ~,
\ee
where the first term is the entropic elastic energy associated
with swelling a polymer chain to a radius $R$, proportional to the
effective spring constant of an ideal chain, $k_BT / R_0^2$,
and the second term
is the second-virial repulsive energy  proportional to the
coefficient $v_2$,  and the segment density squared. It is
integrated over the volume $R^3$. The optimal radius $R$ is
calculated by minimizing  this free energy and gives the swollen
radius
\be \label{swollenflex} R_F \sim a (v_2/a^3)^{1/5} N^\nu ~,
\ee
with $\nu = 3/5$. For purely steric interactions with $v_2 \simeq
a^3$ we obtain $R \sim a N^\nu$. For weakly interacting monomers,
$0< v_2  <a^3$, one finds that the
swollen radius Eq.(\ref{swollenflex}) is only realized above a
minimal monomer number $N \simeq (v_2/a^3)^{-2}$ below which
the chain statistics is unperturbed by the interaction and the
scaling of the chain radius is Gaussian and given by
Eq.(\ref{idealscaling}). 

In the opposite limit of a negative second virial coefficient, $v_2 < 0$,
corresponding to the bad or poor solvent regime,
the polymer coil will be collapsed due to attraction between monomers.
In this case, the attraction term in the free energy
is balanced by the third-virial term in a
low-density expansion (where we assume that $v_3 >0$),
\be
F \simeq v_2 R^3 \left( \frac{N }{R^3} \right)^2
+v_3 R^3 \left( \frac{N }{R^3} \right)^3~.
\ee
Minimizing this free energy with respect
to the chain radius one obtains
\be
\label{collapsed} R \simeq (v_3 / |v_2| )^{1/3} N^\nu
\ee
with $\nu = 1/3$. This indicates the formation of a compact
globule, since the monomer density  inside the globule, $\rho  \sim
N / R^3 $, is independent of the chain length. The minimal  chain
length  to observe a collapse behavior is $N \sim (v_3
/ a^3 v_2)^2$.  For not too long chains and a second virial
coefficient not too much differing from zero, the interaction
is irrelevant and one obtains effective Gaussian or
ideal behavior. It should be noted, however, that even small
deviations from the exact theta conditions (defined by strictly
$v_2 =0$) will lead to chain collapse  or swelling for very long
chains.

\subsection{Charged polymers}

A polyelectrolyte (PE) is a polymer with a fraction $f$ of charged
monomers. When this fraction is small, $f \ll 1$, the
PE is weakly charged, whereas when $f$ is close to
unity, the polyelectrolyte is strongly charged. There are two
common ways to control $f$~\cite{barrat1}. One
way is to polymerize a
heteropolymer mixing  strongly acidic and neutral monomers as building
blocks. Upon contact
with water, the acidic groups dissociate into positively charged
protons (H$^+$) that bind immediately to water molecules, and negatively
charged monomers. Although this process effectively charges the
polymer molecules, the counter-ions make the PE solution
electro-neutral on larger length scales.
The charge distribution along the chain is quenched
 during the polymerization stage, and it is
characterized by the fraction of charged monomers on the chain,
$f$. In the second way, the PE is a weak polyacid or
polybase. The effective charge of each monomer is controlled by
the $pH$ and the salt concentration of the solution\cite{Netzanneal1}. 
Moreover, this annealed fraction depends
on the local electric potential which  is in particular important
for adsorption or binding processes since the local electric potential close to a
strongly charged surface or a second 
charged polymer can be very different from its value in
the bulk solution and therefore modify the polyelectrolyte charge\cite{Netzanneal2}.

Counterions are attracted to the charged polymers via
long-ranged Coulomb interactions; this physical association
typically leads to a rather loosely bound counter-ion cloud
around the PE chain.
Because of this background of a polarizable and
diffusive counterion cloud, there is a strong influence of the
counterion distribution on the PE structure and vice versa. 
Counterions contribute significantly
towards bulk properties, such as the osmotic pressure, and their
translational entropy is responsible for the generally good water
solubility of charged polymers. In addition, the statistics of PE
chain conformations is governed by intra-chain Coulombic repulsion
between charged monomers; this results in a more extended and
swollen conformation of PE's as compared to neutral polymers 
and gives rise to the characteristically high viscosity of polyelectrolyte
solutions (hence their use as viscosifiers in food industry).

For polyelectrolytes, electrostatic interactions provide the driving force
for their salient features and have to be included in
any theoretical description. The reduced electrostatic interaction
between two point-like charges
 can be written as $ q_1 q_2 v(r)$ where
\begin{equation} \label{intro1}
v (r) = \ell_B / r
\end{equation}
is the Coulomb interaction between two elementary charges
in units of $k_BT$ and $q_1$
and $q_2$ are the valencies (or the reduced charges in units of
the elementary charge $e$).
The Bjerrum length $\ell_B$ is defined as
\begin{equation}
\ell_B = \frac{e^2}{4 \pi  \varepsilon k_BT},
\end{equation}
where $\varepsilon$ is the medium dielectric constant.
It denotes the distance at which the Coulombic interaction
between two unit charges in a dielectric medium is equal to
thermal energy ($k_BT$). It is a measure of the distance below
which the Coulomb energy is strong enough to compete with the
thermal fluctuations; in water at room temperatures, one finds
$\ell_B \approx 0.7$\,nm.

 In  biological systems and most industrial applications, the
aqueous solution contains in addition to the counterions
mobile salt ions. Salt ions of opposite
charge are drawn to the charged object and modify the 
counterion cloud around it. They effectively reduce or {\em
screen} the charge of the object. The effective (screened)
electrostatic interaction between two charges $q_1$ and $q_2$
in the presence of salt ions and a polarizable solvent can be
written as $q_1 q_2 v_{\rm DH}(r)$, with the Debye-H\"uckel (DH)
potential $v_{\rm DH}(r)$ given (in units of $k_BT$) by
\begin{equation}
 \label{introDH}
v_{\rm DH} (r) = \frac{\ell_B}{r} {\rm e}^{-\kappa r}~.
\end{equation}
 The exponential decay is characterized
by the screening length $\kappa^{-1}$, which is related to the
salt concentration $c_\s$ by
$\kappa^2 = 8 \pi q^2 \ell_B c_\s~$,
where $q$ denotes the valency of $q:q$ salt. At
physiological conditions the salt concentration is  $c_\s \approx
0.1$\,M and for monovalent ions ($q=1$) this leads to $\kappa^{-1}
\approx 1$\,nm. This means that although the Coulombic interactions
are long-ranged, in physiological conditions they are highly screened
above length scales of a few nanometers, which results from multi-body
correlations between ions in a salt solution.

 For charged polymers, the effective bending stiffness and thus the Kuhn length
is increased due to electrostatic repulsion between monomers\cite{Odijk0,Skolnick,barrat2,Netz2,Netz2b,Sim3,Sim4}. This
effect modifies considerably not only the PE behavior in solution
but also their adsorption characteristics\cite{Netz6}.

A peculiar phenomenon occurs for highly charged PE's and is known
as the Manning condensation of counter-ions~\cite{Man1,Man2,Man3}.
This phenomenon constitutes a true
phase transition in the absence
of added salt ions.
For a single  rigid PE chain represented by an infinitely
long and straight cylinder
with a linear charge density $\tau$  larger than the threshold
\be \label{Manning}
\ell_B \tau q =1~,
\ee
where $q$ is the counter-ion valency, it was shown that
counter-ions condense on the oppositely charged cylinder even in
the limit of infinite system size. For a solution
of stiff charged polymers this corresponds to  the limit
where the inter-chain distance tends to infinity. This effect
is not captured by the linear Debye-H\"uckel theory. 
A simple heuristic way to incorporate the
non-linear  Manning condensation is to replace the bare
linear charge density $\tau$ by the renormalized one, $\tau_{\rm
renorm} = 1/(q \ell_B)$, whenever $\ell_B \tau q
>1$ holds. This procedure, however, is not totally
satisfactory at high salt concentrations
\cite{Fixman2,LeBret}. Also, real polymers have a finite length,
and are neither completely straight nor in the infinite dilution
limit \cite{Man4}. Still, Manning condensation has an
experimental significance for polymer solutions\cite{Man5}
because thermodynamic quantities, such as counter-ion
activities~\cite{Wandrey} and osmotic coefficients~\cite{Blaul},
show a pronounced signature of Manning condensation. Locally,
polymer segments can be considered as straight over length scales
comparable to the persistence length. The Manning condition
Eq.~(\ref{Manning}) usually denotes a region where the binding of
counter-ions to charged chain sections begins to deplete the
solution from free counter-ions. 

\section{Neutral grafted polymers}
\subsection{Scaling approach}

The scaling behavior of the brush height $h$ can be analyzed
using a Flory-like mean--field theory, which is a simplified
version of the original Alexander theory~\cite{alex} for polymer brushes. The
stretching of the chain leads to an entropic free energy loss of
$h^2/R_0^2$ per chain, and the repulsive energy density due to
unfavorable monomer-monomer contacts is proportional to the
squared monomer density times the excluded-volume
parameter $v_2$. The derivation is thus analogous to the calculation
of the Flory radius of a chain in good solvent shown in Section 2.1, 
except that now the grafting density $\rho_a$ plays a decisive role and
controls the amount of stretching of the chains.
The free energy per
chain (and in units of $k_B T$) is then
\be
\label{flory}
 F \simeq \frac{h^2}{a^2 N} +
v_2 \left(\frac{\rho_a N}{h}
\right)^2 \frac{h}{\rho_a}~.
\ee
The equilibrium height is obtained by minimizing  Eq.~(\ref{flory})
with respect to $h$, and the result is~

\be
\label{Floryh} h_0 = N \left( 2 v_2 a^2 \rho_a /3 \right)^{1/3}~ \ee
where the numerical
constants have been added for numerical convenience
in the following considerations.
The vertical size of the brush scales linearly with the
polymerization index $N$, a clear signature of the strong
stretching of the polymer chains, as was originally obtained by Alexander
\cite{alex}. At the overlap threshold,
$\rho_a^* \sim a^{-2} N^{-6/5}$, the height scales as $h_0 \sim N^{3/5}$,
and thus agrees with the scaling of an unperturbed chain radius in
a good solvent, Eq.(\ref{swollenflex}), as it should.
The simple scaling calculation
predicts the brush height $h$ correctly in the asymptotic limit of
long chains and strong overlap. It has been confirmed by
experiments~\cite{auroy1,taunton,field} and computer
simulations~\cite{cos87,murat}.

\subsection{Mean-field calculation}

The above scaling result assumes that all chains are stretched to
exactly the same height, leading to a step-like shape of the
density profile. Monte-Carlo and numerical mean--field calculations
confirm the general scaling of the brush height, but exhibit a
more rounded monomer density profile which goes continuously to
zero at the outer perimeter~\cite{cos87}.
A big step towards a better understanding of stretched polymer
systems was made by Semenov~\cite{sem}, who recognized the
importance of {\em classical paths} for such systems.

The classical polymer path is defined as the path which minimizes
the free energy, for given start and end positions, and thus
corresponds to the most likely path a polymer can take. The name
follows from the analogy with quantum mechanics, where the
classical motion of a particle is given by the quantum path with
maximal probability. Since for strongly stretched polymers the
fluctuations around the classical path are weak, it is expected
that a theory that takes into account only classical paths, is a
good approximation in the strong-stretching limit. To quantify the
stretching of the brush, let us introduce the (dimensionless)
interaction  parameter $\beta$, defined as
\be
\beta \equiv N\left(\frac{3 v_2^2 \rho_a^2 }{2 a^2 }\right)^{1/3}
=\frac{3}{2} \left( \frac{h_0}{a N^{1/2}} \right)^2~,
\ee
where $h_0$ is the
brush height according to Alexander's theory, compare
Eq.~(\ref{Floryh}). The parameter $\beta$ is proportional to the
square of the ratio of the Alexander prediction for the brush
height, $h_0$, and the unperturbed Gaussian chain radius $R_0 \sim a N^{1/2}$,
and, therefore, is a measure of the stretching of the brush based
on the scaling prediction. We will later see that the actual stretching
of the brush is not correctly described by the scaling result for short or
weakly interacting chains. 
Constructing a classical theory in the infinite-stretching limit,
defined as the limit $\beta \rightarrow \infty$, it was shown
independently by Milner et al.~\cite{mil} and Skvortsov et
al. \cite{skvor}
that the resulting monomer density profile $\rho(z)$
depends only on the vertical distance $z$ from the grafting surface
and  has in fact a {\em parabolic} profile. Normalized
to unity, the density profile is given by\cite{mil,skvor}
\be \label{paraprofile}
\frac{\rho(z) h_0} {\rho_a N} =  \left(\frac{3 \pi}{4}\right)^{2/3} -
\left(\frac{\pi z }{2 h_0}\right)^2~.
\ee

\noindent The brush height, \ie, the value of $z$ for which the
monomer density becomes zero, is given by  $z^* = (6/\pi^2)^{1/3}
h_0$ and is thus proportional to the scaling prediction for the brush
height, Eq.(\ref{Floryh}).
The parabolic brush profile has subsequently been confirmed in
computer simulations~\cite{cos87,murat} and
experiments~\cite{auroy1} as the limiting density profile in the
strong-stretching limit, and constitutes one of the cornerstones
in this field. Intimately connected with the density profile is
the distribution of {\em polymer end points}, which is non-zero
everywhere inside the brush (as we will demonstrate later on), 
 in contrast with the original scaling
description leading to Eq.~(\ref{Floryh}).

However, deviations from the parabolic profile become
progressively important as the length of the polymers $N$ or the
grafting density $\rho_a$ decreases. In a systematic derivation of
the mean--field theory for  Gaussian brushes~\cite{netzbrush} it
was shown that the mean--field theory is  characterized  by a
single parameter, namely the stretching parameter $\beta$. In the
limit $\beta \rightarrow \infty$, the difference between the
classical approximation and the mean--field theory vanishes, and
one obtains the parabolic density profile. For finite $\beta$ the
full mean--field theory and the classical approximation lead to
different results and  both show deviations from the parabolic
profile.

In Fig.3 we show the density profiles
(normalized to unity)
for four different values
of $\beta$, obtained with the full mean--field
theory~\cite{netzbrush}. 
In a) the distance from the grafting surface is rescaled  by the 
scaling prediction for the brush height, $h_0$, and in b) it is
rescaled by the unperturbed polymer radius $R_0$.
For comparison, we also show the asymptotic result according to
Eq.~(\ref{paraprofile}) as  dashed lines.  The
self-consistent mean--field equations are solved in the continuum
limit, where the results depend only on the single
parameter  $\beta$ and direct comparison with other continuum
theories becomes possible.
As $\beta$ increases, the density profiles approach the
parabolic profile  and
already for $\beta = 100$ the
density profile obtained within mean--field theory is almost
indistinguishable from the parabolic profile denoted by a thick
dashed line in Fig. 3a. What is interesting to see is that for small values
of the interaction parameter $\beta$, the numerically determined
density profiles exhibit a larger brush height than the asymptotic prediction.
This has to do with the fact that entropic effects, due
to steric polymer repulsion from the grafting surface, are not accounted
for in the infinite-stretching approximation.
Experimentally, the achievable $\beta$ values are 
below $\beta \simeq 50$, which means that deviations from the
asymptotic parabolic profile are important. For moderately large
values of $\beta >10$, the classical approximation (not shown
here), derived from the mean--field theory by taking into account
only one polymer path per end-point position, is still a good
approximation, as judged by comparing density profiles obtained
from both theories~\cite{netzbrush}, except very close to the
surface. Unlike mean-field theory, the classical theory
misses completely the depletion
effects at the substrate. Depletion effects at the substrate lead to a
pronounced density depression close to the  grafting surface, as
is clearly visible in Fig.~3.

Let us now turn to the thermodynamic behavior of a polymer brush.
Using the Alexander scaling model, we can calculate the
free energy per chain
by putting the result for the optimal brush height, Eq.~(\ref{Floryh}),
into the free-energy expression, Eq.~(\ref{flory}), and obtain
\be \label{freealex}
 F \sim \beta \sim N \left( v_2 \rho_a / a \right)^{2/3}.
\ee
In Figure 4a we show the rescaled free energy per chain 
within the full mean-field framework (solid line) in comparison
with the infinite stretching result (dotted horizontal line) and including
the leading correction due to the finite entropy of the end-point distribution
(broken line)\cite{netzbrush}. 
In the infinite-stretching limit, i.e. for $1/\beta  \rightarrow 0$,
all curves converge. The free energy is not directly measurable, but
the lateral pressure can be determined using the Langmuir film balance technique.
The osmotic surface pressure $\Pi$
is related to $F$, the free energy per chain, by
\be
\Pi = -\frac{\partial (F \rho_a A)}{\partial A} = 
\rho_a^2 \frac{\partial F}{\partial \rho_a} =
\frac{2 \rho_a }{3} \left( F - \frac{\partial F /\beta }{\partial \beta^{-1}} \right)
\ee
In Figure 4b we show the rescaled osmotic pressure obtained within the mean-field
approach (solid line). In  the infinite-stretching limit, one expects a scaling
as $\Pi \sim \rho_a^{5/3}$  which is shown as a dotted line. One notes that the
more accurate mean-field result gives a pressure which is strictly larger than 
the asymptotic infinite stretching result, in agreement with previous calculations and
experiments\cite{carignano,kent,martin,baran,currie}. 
In the presence of excluded-volume correlations, \ie,
when the chain overlap is rather moderate, the scaling of the
brush height $h$, Eq.(\ref{Floryh}), 
is still correctly predicted by the Alexander calculation, but the
prediction for the free energy, Eq.(\ref{freealex}), 
is in error. Including correlations
\cite{alex}, the free energy is predicted to scale as $ F \sim N \rho_a^{5/6}$. 
leading to a pressure which scales as 
$\Pi \sim \rho_a^{11/6}$
in the presence of correlations. However, all
these theoretical predictions do not compare well
with experimental results for the surface
pressure of a compressed brush~\cite{kent}
which has to do with the fact that experimentally, chains are rather short so that
one essentially is dealing with a crossover situation between the mushroom
and brush regimes.
An alternative theoretical method to study tethered chains is the so-called
single-chain mean--field method~\cite{carignano}, where the statistical
mechanics of a single chain is treated exactly, and the interactions
with the other chains are  taken into account on a mean-field level.
This method is especially useful for short chains, where
fluctuation effects are important, and for dense systems,
where excluded volume interactions play a role. The calculated
profiles and brush heights agree very well with experiments and
computer simulations. Moreover, these calculations explain the pressure isotherms
measured experimentally~\cite{kent} and in molecular-dynamics
simulations~\cite{grest}.

A further interesting question concerns the behavior of individual
polymer paths. As was already discussed for the infinite-stretching
theories ($\beta \rightarrow \infty$),  polymers paths
do end at any distance from the surface. 
In the left part of Figure 5 we show
mean-field  results for the rescaled averaged polymer 
paths which end at a certain distance $z_e$ from the wall
for $\beta =\,\,$1, 10, 100 (from top to bottom). Analyzing the polymer paths
which end at a common distance from the surface, two rather
unexpected features are obtained: i) free polymer ends,  in
general, are stretched; and, ii) the end-points lying close to the
substrate are pointing towards the surface
 (such that the polymer path first
turns away from the grafting surface before moving back
towards it). In contrast, end-points lying
 {\it beyond} a certain distance from the
substrate, point away from the surface
(such that the paths move monotonously towards the
surface).
As we will explain shortly below,  these two features have been
confirmed in molecular-dynamics simulations~\cite{Seidel}.
They are not an artifact of the continuous self-consistent theory
used in Ref.~\cite{netzbrush} nor are they due to the neglect of
fluctuations. These are interesting results, especially since it
has been long assumed that free polymer ends are
unstretched, based on the assumption that no
forces act on free polymer ends.
The thick solid line shows 
the unconstrained mean path obtained by averaging over all end-point 
positions. Note that the end-point stretching is small but finite for all finite
stretching parameters $\beta$. The right panel exhibits the end-point distributions,
which, obviously, are finite over the whole range of vertical heights.

\subsection{Molecular Dynamics simulations}

We now present results from Molecular Dynamics simulations
in which all the chain monomers are coupled to a
heat bath. The chains interact via the repulsive portion of a 
shifted  Lennard-Jones potential with a Lennard-Jones diameter
$\sigma$, which corresponds to
a  good solvent situation.
For the bond potential between adjacent polymer segments we take a  
FENE (nonlinear bond) potential which gives an average nearest-neighbor
monomer-monomer separation of typically $a \approx 0.97 \sigma$.
In the simulation box with a  volume $L\times
L\times L_z$ there are 50 (if not stated otherwise)
 chains each of which consists of $N+1$
monomers with varying monomer number $N=20,\,30,\,50$, as indicated.
 The box length $L$ in
$x-$ and  $y-$direction was chosen to give anchoring densities
$\rho_a$ from 0.02$\,\sigma^{-2}$ to 0.17$\,\sigma^{-2}$.
The first monomer of each chain  is firmly
and randomly attached to the grafting surface at $z=0$. 
The mean square end-to-end distance of identical free (i.e. unanchored) 
chains was for the case $N=50$ found to  be 
$R_F \equiv \langle({\bf r}_N - {\bf  r}_0)^2\rangle_{\rm free}^{1/2}=12.55\sigma$. 
For more details consult \cite{Seidel}.

Figure~6 shows the behavior of the reduced monomer
density $\rho(z) R_F /N\rho_a\,$ at increasing anchoring density. The
stretching of the chains with increasing surface coverage, which
is due to the repulsion between monomers, is evident. This plot
has to be compared with Figure 3b, where the same type of rescaling
has been used. However, note that at this point, direct and quantitative
comparison is not possible, since it is a priori not clear
which value of the interaction parameter $\beta$ in the self-consistent
calculation corresponds to which set of simulation parameters
$\sigma$, $N$, $\rho_a$.

The theoretically predicted scaling law of the brush height $h_0\sim
N\rho_a^{1/3} $ has been confirmed in several simulations\cite{murat}. 
Provided $\rho_a$ is
above the critical overlap density $\rho_a^*\sim N^{-6/5}$, the brush
height, as measured by the first moment of the monomer density
distribution $\langle z\rangle  =  \int\!  z  \rho(z) dz/\! \int\!\rho(z) dz$, 
should approach the predicted scaling form.
Figure 7 shows this behavior in an appropriate scaling
plot of the average brush height $\langle z \rangle $
and the $z$-component of the radius of
gyration $R_G \equiv \langle\, \sum_{i=0}^N({\bf r}_i - {\bf
  r}_{cm})^2 /(N+1) \rangle^{1/2} $ where ${\bf  r}_{cm}$ is the position of
the center of mass.
For the scaling plot the brush height and radius of gyration are
divided by the scaling prediction $h_0 \sim N \rho_a^{1/3} a^{5/3} $
and plotted as a function of the analogue of the interaction 
parameter, namely $h_0 / R_F \sim N \rho_a^{1/3} a^{5/3} / R_F$,
where $a$ and $R_F$ are determined  from the simulation in the
absence of a grafting wall.
In this way we obtain an universal crossover point 
 $N\rho_a^{1/3} a^{5/3} / R_F \approx 1.4$. 
Thus, for the chain length $N$=50, the case we discussed most,
one can be sure to reach the asymptotic scaling regime for the larger 
 grafting densities.
 
A general problem when comparing experimental, simulation and
analytical results among each other is that the different parameters
have to be matched in a meaningful way.  One such way is based
on the relative stretching of chains in a brush. In Figure~8
we plot in a) simulation results for the stretching ratio
\be
\gamma =\,\langle z_e({\rho_a})\rangle/\langle
z_e({\rho_a}=0)\rangle
\ee
which is the averaged end-point height for a finite grafting
density divided by the end-point height for vanishing grafting density.
The stretching ratio is plotted as a function of 
the scaled anchoring density 
$(N\rho_a^{1/3} a^{5/3} /R_F)^2$ which 
is the analogue of the interaction parameter $\beta$.  
It is seen that the different chain lengths and grafting densities
scale quite nicely. Clearly, for short chains or low grafting densities, 
that is in the mushroom regime,
the stretching ratio $\gamma$ approaches unity. The highest
stretching ratio reached in the simulations  is $\gamma \approx 2.7$.
In Figure 8b we show mean-field results for the stretching ratio
$\gamma$ as a function of the interaction parameter $\beta$.
The general shape of the curve is similar. Matching the 
stretching ratios of the simulation and the mean-field calculation
allows to determine the effective $\beta$ value corresponding to
a particular simulation run. As an example, for the most highly
stretched simulation with $\gamma \approx 2.7$, we 
estimate an effective interaction parameter of $\beta \approx 25$. 
A similar matching procedure is possible with experimental data.

In agreement with the mean-field  profiles shown in
Figure 3,  the monomer density obtained from the simulations
in Figure 6 decays smoothly to zero at distances far from the anchoring plane. 
However, 
comparing the shape of the profiles in more detail, one notes 
 a region over which the simulation  profiles become rather flat for anchoring
densities $\rho_a > 0.1\sigma^{-2}$, in contrast to the
mean-field results. According to
Figure~8, this corresponds to stretching factors 
$\gamma > 2$ or $\beta > 10$. 
 This discrepancy can be traced back to the neglect of higher-order virial terms in
 the analytical theory, which beome important at elevated monomer densities.
 
 We also looked at individual polymer paths within the simulation.
 The scaled average polymer paths with a given fixed 
end-point distance $\,z_e$,
 $\langle z_{z_e}(n)\rangle / \langle z_e\rangle$,
are shown in Figure~9 for three typical anchoring densities 
corresponding to stretching factors $\gamma \approx\,\,$1.3, 1.8, and
2.7, from top to bottom. To get an idea of their relative weight we plot the
polymer paths together with the normalized density of
free ends $\,\rho_e(z)\,$ (shown in the right panel).   
The trajectories are very similar to those predicted by the mean-field 
theory (see Figure~5). Paths which end far from
the anchoring surface are stretched through their entire length,
including the free end point. The paths 
which end at the outer rim of the distribution are almost uniformly stretched and appear
almost as a straight line. On the other hand, paths which end close to the
anchoring surface are non-monotonic, first moving away from the wall, reaching
a maximal distance, and then turning back towards the anchoring surface. Except at the
maximum, all paths are stretched everywhere, including the end point. 
 The straight broken lines are added for comparison and
denote the maximally stretched possible path,
 which consists of a fully oriented and straight polymer.
As can be seen from
Figure~5 as well as from Figure~9, the end-point 
stretching, proportional to $(dz/ds)$ or $(dz/dn)$, respectively, is positive
for  some particular paths and negative for others, so that the average,
plotted as thick line in the Figures, is typically quite small.
 This reconciles the present results with the infinite-stretching results by
Milner, Witten and Cates
\cite{mil} and Zhulina and co-workers
\cite{skvor}, where the chains were assumed to be unstretched at their ends
regardless of their end-point position: this assumption turns out to be
acceptable on average if the stretching of the chains is large, i.e., for 
large $\beta$.

\subsection{Additional effects}
As we described earlier, the main interest in end-adsorbed or
grafted polymer  layers stems from their ability to stabilize
surfaces against van-der-Waals attraction. The force between
colloids with grafted polymers is repulsive if the polymers do not
adsorb on the grafting substrates~\cite{collgraft}, that is, in the absence of
 polymer bridges and loops. A stringent test of brush theories was
possible with accurate experimental measurements of the repulsive
interaction between two opposing grafted polymer layers using a
surface force apparatus~\cite{taunton}. The resultant force could
be fitted very nicely by the infinite-stretching theory of Milner
et al.~\cite{milner88b}. It was also shown that polydispersity
effects, as appear in experiments, have to be taken
into account theoretically in order to obtain a good fit of the
data~\cite{milner89}.

So far we assumed that the polymer grafted layer is in contact
with a good solvent. In this case, the grafted polymers try to
minimize their mutual contacts by stretching out into the solvent. If the
solvent is bad, the monomers try to avoid the solvent by forming a
collapsed brush, the height of which is considerably reduced with
respect to the good-solvent case. It turns out that the collapse
transition, which leads to phase separation in the bulk, is
smeared out for the grafted layer and does not correspond to a
true phase transition~\cite{halperin88}. The height of the
collapsed layer scales linearly in $\rho_a N$, which reflects the
constant density within the brush, in agreement with
experiments~\cite{auroy2}. Some interesting effects have been
described theoretically~\cite{marko93} and
experimentally~\cite{auroy2} for brushes in mixtures of good and
bad solvent, which can be rationalized in terms of a partial
solvent demixing.

For a theta solvent ($v_2=0$) the relevant interaction is
described by the third-virial coefficient; using a simple
Alexander approach similar to the one leading to
Eq.~(\ref{Floryh}), the brush height is predicted to vary with the
grafting density as $ h \sim \rho_a^{1/2}$, in agreement with
computer simulations~\cite{theta}.

Up to now we discussed planar grafting layers. It is of much
interest to consider the case where
polymers are grafted to {\it curved} surfaces. The first study taking
into account curvature effects of stretched and tethered polymers
was done in the context of star polymers~\cite{daoudcotton}. It
was found that chain tethering in the spherical geometry leads to
a universal density profile, showing a densely packed core, an
intermediate region where correlation effects are negligible and
the density decays as $\rho(r) \sim 1/r$, and an outside region
where correlations are important and the density decays as $\rho
\sim r^{-4/3}$, with the radial distance denoted by $r$. 
These considerations were  extended using  the
infinite-stretching theory of Milner et al.~\cite{ball},
self-consistent mean--field theories~\cite{dan},  and
molecular-dynamics simulations~\cite{murat91}. Of particular
interest is the behavior of the bending rigidity of a polymer
brush, which can be calculated from the free energy of a
cylindrical and a spherical brush and forms a conceptually simple
model for the bending rigidity of a lipid
bilayer~\cite{milnerbend}.

The behavior of a polymer brush in contact with a polymeric solvent, 
consisting of chemically identical
but somewhat shorter chains than the brush, had been first
considered by de Gennes~\cite{gennes}. A complete scaling
description has been given only recently~\cite{aubouy}. One
distinguishes different regimes where the polymer solvent is
expelled to various degrees from the brush. A somewhat related
question concerns the behavior of two opposing brushes brought
closely together, and
separated by a
solvent  consisting of a polymer solution~\cite{gast,boru}. Here one
distinguishes a regime where the polymer solution leads to a
strong attraction between the surfaces via the ordinary depletion
interaction, but also a high
polymer concentration regime where the attraction is not  strong
enough to induce colloidal flocculation. This phenomenon is called
colloidal restabilization~\cite{gast}.

Considering a mixed brush made of
mutually incompatible grafted chains, a novel transition to a
brush characterized by a lateral composition modulation was
found~\cite{marko91}. Even more complicated spatial structures are
obtained with grafted diblock copolymers~\cite{brown}. Finally, we
would like to mention in passing that  these static brush
phenomena have interesting consequences for the dynamic properties of
polymer brushes~\cite{halperin88b}.

\section{Charged grafted polymers}

Brushes can also be formed by
charged polymers which are densely end-grafted to a surface;
the resulting  {\em charged brush} shares many of the features discussed above
for neutral brushes, but qualitatively new properties emerge
due to the presence of charged monomers and counterions in the brush. 
Charged brushes have
been the focus of numerous theoretical
~\cite{MIK88}-\cite{ZHU97} and experimental
~\cite{MIR95}-\cite{balast} studies. 
They serve as an efficient mean for preventing colloids
in polar media (such as aqueous solutions) from flocculating and
precipitating out of solution. This stabilization
arises from steric (entropic) as well as electrostatic repulsion.
A strongly charged brush is able to trap its own counter-ions and
generates a layer of locally enhanced salt
concentration~\cite{PIN91}. It is thus less sensitive to the
salinity of the surrounding aqueous medium than a stabilization
mechanism based on pure electrostatics (\ie\  without polymers).
Compared to the experimental
knowledge about uncharged polymer brushes,
less is known  about the
scaling behavior of PE brushes.
The thickness of the brush layer
has been inferred from neutron-scattering experiments on
end-grafted polymers ~\cite{MIR95} and charged
diblock-copolymers at the air-water interface ~\cite{AHR97}.

Theoretical work on PE brushes was initiated by the works of
Miklavic and Marcelja ~\cite{MIK88} and Misra et al.
~\cite{MIS89}. In 1991, Pincus ~\cite{PIN91} and Borisov,
Birshtein and Zhulina ~\cite{BOR91} presented scaling theories for
charged brushes in the so-called osmotic regime, where the brush
height results from the balance between the chain elasticity
(which tends to decrease the brush height) and the repulsive
osmotic counter-ion pressure (which tends to increase the brush
height). In later studies, these works have been generalized to
poor solvents ~\cite{ROS92} and to the regime where excluded
volume effects become important, the so-called quasi-neutral or
Alexander regime~\cite{BOR94}.

\subsection{Scaling approach}

An analytical theory for polyelectrolyte brushes relies on a number of simplifying assumptions. 
The full theoretical problem is intractable because the degrees of freedom of the 
polymer chains and the counterions are coupled by steric and long-ranged Coulomb 
interactions. 
It is important to note that charged polymers by itself are not fully 
understood, therefore quite drastic simplification are needed 
to tackle the more complicated system of polyelectrolytes 
end-grafted to a surface. 
Firstly, we will concentrate on polymeric systems with counterions and
only briefly mention the effects  of added salt towards the end of this section, 
which has been discussed extensively in the
original literature. 
Secondly, we will write the total free energy per unit area and
in units of $k_BT$, 
\be F = F_{\rm pol}+F_{\rm ion} + F_{\rm int}
\ee
as a sum of separate contributions from the polyelectrolytes, $F_{\rm pol}$,
contributions from the counterions, $F_{\rm ion}$,
and an electrostatic interaction term $F_{\rm int}$ which couples polymers and counterions.
The schematic geometry of the brush system is visualized in Figure 10: 
we assume that the charged brush is characterized
by two length scales.
The polymer chains are assumed to extend to a distance $h$ 
from the grafting surface, 
the counterions in general form a layer with a thickness of $d$. 
Two different scenarios emerge: The
counter-ions can either extend outside the brush, $d \gg h$, as
shown in Figure 10a, or be confined inside the brush, $d\approx h$ as
shown in Figure 10b. As we demonstrate now, case b) is indicative of strongly
charged brushes, and thus applicable to most experiments done on charged brushes so far,
 while case a) is typical for weakly charged brushes.

 We recall that
the grafting density of PE's is denoted by $\rho_a$, $q$ is the
counter-ion valency, $N$ the polymerization index of grafted
chains, and $f$ the charge fraction.
 The counterion free energy contains entropic contributions (due to the confinement of the 
counterions inside a layer of thickness $d$) and also energetic contributions which come 
from interactions between counterions. In previous theories, a low-density expansion for 
the interaction part was used. We remind that the second-virial interaction is important for 
neutral brushes and is the driving repulsive force balancing the chain elasticity. 
For charged brushes, on the other hand,
the leading term of the electrostatic correlation energy (which 
shows fractional scaling with respect to the charge density) is attractive and has been 
shown to lead to a collapse of the polyelectrolyte brush for large 
Bjerrum lengths\cite{Csajka,Lau}. We will not pursue this collapse transition
further in this review paper but solely concentrate on entropic and steric counterion effects,
which is an acceptable approximation for mono-valent counterions..
In the present analysis we use a free-volume approximation very much in the spirit of 
the van-der-Waals equation for the liquid-gas transition. We include 
the effective hard-core volume of a single polyelectrolyte chain, which we call $V$, and 
which reduces the free volume that is available for the counterions. This free volume 
theory therefore takes the hard-core interactions between the polymer monomers 
and the counterions into account in a non-linear fashion and is valid even
at large densities in the limit of close-packing. Compared to that, the 
excluded-volume interaction between counterions is small since the monomers 
are more bulky than the counterions and therefore it is neglected. 
The non-linear entropic free energy contribution of the counterions reads per unit area
\begin{equation} \label{freeion}
 F_{\rm ion}  \simeq
\frac{N f \rho_a}{q} \left[ 
       \ln \left( \frac { N f \rho_a/q}{ d-\rho_a V }\right) - 1 \right]~.
\end{equation}
In the limit of vanishing polymer excluded-volume, $V\rightarrow 0$, 
one recovers the standard ideal entropy expression. 
As the volume available for the counterions in the brush, 
which per polymer is just $d/\rho_a$, 
approaches the self volume of the polymers, $V$, 
the free energy expression Eq. (\ref{freeion}) diverges, 
that means, the entropic prize for that scenario becomes infinitely large. 
The excluded volume of the polymers is roughly independent of the polymer 
brush height, and can be written in terms of the effective monomer  
hardcore diameter $\sigma_{\rm eff} $  
and the polymer contour length $aN$ as $V \approx aN \sigma_{\rm eff} ^2$. 
This leads to the final expression 
\begin{equation} \label{freeion2}
 F_{\rm ion}  \simeq
\frac{N f \rho_a}{q} \left[ 
       \ln \left( \frac { N f \rho_a/q}{ d-N a \rho_a \sigma_{\rm eff} ^2 }\right) - 1 \right]~.
\end{equation}
The polymer free energy $F_{\rm pol}$ per unit area
can in principle be derived from the free energy per chain  used for neutral brushes,
Equation (\ref{flory}), by multiplying with the grafting density. It contains 
two contributions, one  due to the chain elasticity and the other due to monomer-monomer
interactions. Note that a logarithmic entropy term as for the counterions is missing, since 
the translational degrees of freedom of the monomers are lost due to chain connectivity.
We recall that the entropy of chain stretching was accounted for by linear elasticity 
theory, with a spring constant proportional to $1/R_0^2$ in units of the thermal energy.
This linear expression for the chain stretching is reliable for the neutral case, where
indeed chain stretching is typically quite mild. For charged chains, the stretching is much 
stronger and we have to consider a more detailed model for the chain elasticity. 
For a freely jointed chain, which is a good model for synthetic polymer chains,
the entropy loss due to stretching can be calculated exactly\cite{ali}. 
We only need here the asymptotic expressions for weak and for strong stretching, 
which read (per unit area)
\begin{equation} \label{pol}
F_{\rm pol}^{\rm el}  \simeq 
     \left\{
\begin{array}{llll}
     & 3 \rho_a h^2/(2 N a^2)
         & {\rm for} &  h \ll Na  \\
     & -\rho_a N \ln (1-h/ Na)
         & {\rm for }     &  h \rightarrow Na \\
\end{array} \right.
\end{equation}
and are proportional to the grafting density  $\rho_a$. 
The contour length of the fully stretched chain is $Na$. 
The weak-stretching term is the standard term used in previous scaling models. 
For the highly stretched situations encountered in fully charged brushes, 
the strong-stretching term is typically more appropriate and 
leads to a few changes in the results as will be explained further below.
The energetic contribution to the polymer free energy 
can be expressed as the second virial contribution, 
arising from steric repulsion between the monomers
(contributions due to counter ions are neglected). Throughout
this section, the polymers are assumed to be in a good solvent
(positive second virial coefficient $v_2 > 0$). The contribution
thus reads
\begin{equation}
  F_{\rm pol}^{\rm en}   \simeq  \frac{1}{2}  h v_2
\left(
      \frac{N \rho_a}{h}\right)^2~.
\end{equation}
An additional electrostatic component to the polymer interaction term
is typically unimportant since the counterions strongly screen any 
Coulomb interactions\cite{ali}.
Finally, an electrostatic interaction between polymers
and counterions  $F_{\rm int}$
occurs if the PE brush is not locally electro-neutral  throughout
the system, as for example is depicted in Fig.~10a. This
energy is given by
\begin{equation}
F_{\rm int} = \frac{2 \pi \ell_{\rm B}
(N f \rho_a)^2}{3}
                \frac{(d-h)^2}{d}~.
\end{equation}
This situation arises in the limit of low charge, when the
counter-ion density profile extends beyond the brush layer, \ie\
$d> h$.

The different free energy contributions lead, upon
minimization with respect to the two length scales $h$ and $d$,
to different behaviors. Let us first consider the weak charging
limit, \ie\  the situation where the counter-ions leave the brush,
$d > h$.
In this case, minimization of ${ F}_{\rm ion} + { F}_{\rm int}$
with respect to the counter-ion height $d$ in the limiting case
of vanishing  brush height ($h=0$) and monomer volume ($\sigma=0$) leads to
 \be
  d = \frac{3}{2 \pi q  \ell_B N f \rho_a} =3 \lambda_{GC}
 \ee
which has the same scaling as
the so-called Gouy-Chapman length $\lambda_{GC}$.
This length scale is a measure for the average height of
the diffuse layer of
$q$-valent counter-ions adsorbed at a
surface with effective surface charge density $ N f \rho_a$
and has been determined within simulations and  field-theory\cite{andre}
but can also be obtained with more coarse-grained scaling arguments,
as demonstrated here.
Balancing
now the polymer stretching energy ${ F}_{\rm pol}^{\rm el}$ and the
electrostatic energy ${ F}_{\rm int}$ one obtains the so-called
Pincus brush height
\begin{equation}
  h \simeq N^3 \rho_a\, a^2\ell_B f^{2}~,
   \label{Pinc}
\end{equation}
which results from the electrostatic attraction between
the counter-ions and the charged monomers. One notes the peculiar
dependence on the polymerization index $N$.
In the limit of $d\approx h$,
the PE brush can be considered as neutral and the electrostatic
energy vanishes. There are two ways of balancing the remaining
free energy contributions. The first is obtained by comparing the
osmotic energy of counter-ion confinement, ${ F}_{\rm ion}$, in the limit
when $d = h$ and for
vanishing polymer volume, with the
polymer stretching term, ${ F}_{\rm pol}^{\rm el}$, in the weak stretching
limit,  leading to the height
 \be \label{osmo}
h \sim \frac{N a f^{1/2}}{(3q)^{1/2}}~,
 \ee
constituting the linear osmotic brush regime. 
The main assumption here is that all counterions stay strictly localized inside
the brush, which will be tested later by comparison with computer simulations. 

Finally
comparing the second-virial term for the counterion interactions, 
${ F}_{\rm pol}^{\rm en}$, with the
polymer stretching energy in the weak-stretching limit, 
${ F}_{\rm pol}^{el}$, one obtains
  the same scaling behavior as
the neutral brush~\cite{alex,gennes}, compare Eq.(\ref{Floryh}).
Comparing the brush heights in all three regimes we arrive at the
phase diagram shown in Fig.~11. The three scaling
regimes coexist at the characteristic charge fraction
\be
f^{\rm co} \sim \left( \frac{q v_2}{N^2 a^2 \ell_B} \right)^{1/3}~,
\ee
and the characteristic grafting density
\be
\rho_a^{\rm co} \sim \frac{1}{N \ell_B^{1/2} v_2^{1/2}}~.
\ee
%
\subsection{Non-linear effects}

The scaling relations for the brush height and the crossover boundaries
between the various regimes constitute the simplest approach towards
charged brushes.  We pointed out already a few limitations of the presented
results, which have to do with non-linear stretching  and finite-volume 
effects.

Computer simulations provide an excellent mean to test those scaling relations. 
Extensive molecular dynamics simulations have been performed recently for
polyelectrolyte brushes at various grafting densities and charge fractions, 
both at strong and intermediate electrostatic couplings\cite{Csajka0,Seidel2}. 
In these simulations, a freely-jointed bead-chain model is adopted 
for charged polymers end-grafted onto a rigid surface. 
The counterions are explicitly modeled as charged particles 
where both counterions and charged monomers are univalent and 
interact with the bare Coulomb potential Eq. (\ref{intro1}). 
The strength of the Coulomb interaction is controlled by the Bjerrum length  $\ell_B$. 
No additional electrolyte is added. 
The short-range repulsion between all particles 
is modeled by a shifted Lennard-Jones (LJ) potential, characterized
by the hard-core diameter $\sigma$, as was done for the neutral brush 
simulations in Section 3.3.
The simulation box is periodic in lateral directions and finite in the z-direction 
normal to the anchoring surface at z=0. We apply the techniques introduced by 
Lekner and Sperb to account for the long-range nature of the Coulomb 
interactions in a laterally periodic system \cite{Csajka0}. 
To study the system in equilibrium we use stochastic 
molecular simulation at constant temperature. 

Simulated density profiles of monomers and counterions of the system in 
normal direction are shown in Figure 12 for the fully charged brush at 
several grafting densities and for a Coulomb coupling characterized by 
$\ell_B /\sigma =1$, which is close to the coupling of monovalent ionic groups in water. 
As seen, both monomers and counterions follow very 
similar nearly-step-like profiles with uniform densities  inside the brush, 
which increases with grafting density. These data show that the counterions 
are mostly confined in the brush layer and that the electroneutrality condition is 
satisfied locally. 
The simple explanation is that the Gouy-Chapman length is indeed
very small and of the order or  smaller than the monomer diameter.
 One can observe  that the polyelectrolyte chains are stretched up to about 70 \% of 
 their contour length, which is roughly $L \approx N \sigma = 30 \sigma$,
and thus their elastic behavior is far beyond the linear regime. 
 Therefore, within the chosen range of parameters, the simulated brush is in the 
 strong-charging (i.e., all counterions are confined within the brush) 
 and strong-stretching limits; as we will show below, it exhibits a 
 non-linear osmotic scaling behavior, so the linear scaling description of the last
 section, leading to Eq. (\ref{osmo}),  has to be slightly modified. 
 The average height of end-points of the chains, $\langle z_e \rangle$,
  is one of the quantities 
 which can be directly measured in the simulations and is shown in Figure 13 
 together with the predictions of the linear scaling theory, Eq. (\ref{osmo}), denoted
 by (a), and the non-linear scaling predictions,  to be explained further below. 
 It is observed that the simulated brush height (solid circles) varies slowly with the 
 grafting density, contrary to the prediction of the linear  scaling theory, 
 Eq.(\ref{osmo}),  but in agreement with recent experimental results\cite{guenoun,helm}.
 
We now present a  scaling theory that incorporates non-linear elastic and osmotic effects.
In order to bring out the physics most clearly, the derivation will be done in two 
consecutive steps. First,
in the strongly-stretched osmotic brush regime, one chooses the strong-stretching 
version of the chain-stretching entropy in Equation (\ref{pol}) 
and balances it with the counterion entropy, Equation (\ref{freeion2}),
assuming  vanishing polymer self volume, $\sigma =0$, and assuming
equal heights for the brush and counterion layers, $d=h$. 
The result is
 \be \label{osmo2}
h \sim \frac{N a q f}{(1+ q f)}~,
 \ee
which is the large-stretching analogue of Equation (\ref{osmo}). 
The maximal stretching predicted from this equation is obtained for f=1 and 
corresponds to a vertical chain extension corresponding to 
 50 \% of the contour length. For comparison, both 
expressions (\ref{osmo}) and (\ref{osmo2}) are shown in Figure 13 as dashed lines (a) 
and (b) respectively. Still, the overall stretching  is considerably smaller than what
is observed in experiments and simulations, and it transpires that something 
is missing in the above scaling description. It has to do with
the entropic pressure which increases as the volume within the brush 
is progressively more filled up by monomers and counterions. Note that the 
brush height in Equations   (\ref{osmo}) and (\ref{osmo2})  does not depend 
on the grafting density, in vivid contrast to the simulation results displayed in Figure 13.
 
In the non-linear osmotic brush regime we combine the high-stretching
(non-linear) version of the chain elasticity in Equation (\ref{pol}) 
with the non-linear entropic effects of the counterions due to the 
finite volume of the polymer chains, i.e. we choose a finite polymer diameter $\sigma$ 
in  Equation (\ref{freeion2}). The final result for the equilibrium brush height is
\be \label{osmo3}
h \sim \frac{N a (q f+  \sigma_{\rm eff} ^2 \rho_a)}{(1+ q f)}~,
\ee
which in the limit of maximal grafting density, that is close packing,  
$\rho_a \rightarrow 1/\sigma^2_{\rm eff} $, 
reaches the maximal value stretching value, $h \rightarrow Na$, 
as one would expect: Compressing the brush laterally increases the 
vertical height and finally leads to a totally extended chain structure.  
In Figure 13, we compare expression Eq. (\ref{osmo3}) with the simulation results
for the brush height as a function of grafting density. 
Note that we have used $\sigma_{\rm eff}^2 =2 \sigma^2$. 
This choice corresponds to an
approximate two-dimensional square-lattice packing of monomers
and counterions on two interpenetrating sublattices.
As can be seen, the scaling prediction, Eq. (\ref{osmo3}), qualitatively captures 
the slow brush height increase with grafting density, as observed in the simulations.
The deviations between simulation data and the non-linear osmotic
brush prediction are unproblematic since the effective hard-core diameter 
of monomers in the simulation is difficult to determine precisely and might
as well be treated as a fitting parameter. Deviations might also be 
explained by considering 
higher-order effects, such as lateral inhomogeneity of the counterion distribution 
around the brush chains and intermediate-stretching elasticity of the 
chains, that go beyond the present scaling analysis and have been 
considered in Ref.\cite{ali}.

\subsection{Additional effects}
For large values of the charge fraction $f$ and the grafting
density $\rho_a$, and for large Coulomb coupling, 
it has been found numerically that the brush height
does not follow any of the scaling laws discussed
here~\cite{Csajka0}. This has been recently rationalized in terms
of another scaling regime, the collapsed regime. In this regime
one finds that correlation and fluctuation effects, which are
neglected in the discussion in this section, lead to a net
attraction between charged monomers and
counter-ions~\cite{Csajka,Lau}. 

If salt is present in the solution, counterions as well as
co-ions do penetrate into the brush, which leads
to additional screening of the Coulomb repulsion inside
the brush. The amount of this screening, and the stretching
of the polyectrolyte chains, are now also controlled by the bulk
salt concentration. Since the additional salt screening 
weakens the swelling of the brush caused by the 
counterion osmotic pressure, salt leads to a brush contraction
for sufficiently high salt concentration according to 
$ h \sim c_{\rm salt}^{-1/3}$\cite{PIN91,BOR91}.
The threshold salt concentration above which the brush contraction
sets in is given by the salt concentration which equals the 
counterion concentration inside the brush. This means that the higher
the grafting density (and consequently the higher the internal
counterion concentration in the osmotic brush regime), 
the larger the salt concentration
neceassary to see any salt effects at all.

Another way of creating a charged brush is to dissolve a diblock
copolymer consisting of a hydrophobic and a charged block in
water. The diblocks associate to form a hydrophobic core, thereby
minimizing the unfavorable interaction with water, while the
charged blocks form a highly charged corona or brush
\cite{Eisenberg}. The micelle morphology depends on different
parameters. Most importantly, it can be shown that salt acts as a
morphology switch, giving rise to the sequence spherical,
cylindrical, to planar micellar morphology as the salt
concentration is increased \cite{Eisenberg}. Theoretically, this
can be explained by the entropy cost of counter-ion confinement
in the charged corona \cite{Netzmicelles}. The charged corona can
be studied by neutron scattering \cite{Guenoun2} or atomic-force
microscopy \cite{Foerster2} and gives information on the behavior
of highly curved charged brushes.

\section{Concluding Remarks}

In this short review about neutral and charged
polymers that are
terminally grafted with  one end to a surface 
we tried to explain and contrast the most important
theoretical methods used for their understanding and description.
 The theories used for brushes  are quite different from 
 ordinary polymer modelling since the
statistics of the grafted layer depends crucially on the fact that
the chain is not attracted to the surface but is forced to be in
contact to the surface since one of its ends is chemically or
physically bonded to the surface. We review scaling concepts, 
 mean--field theory and simulation techniques that 
 give information on brushes on different levels of detail and accuracy.
 Scaling concepts yield the qualitative features of brushes, i.e. the brush height
 and its free energy as a function of the main parameters and without
 reliable prefactors. Mean-field or self-consistent theories allow to construct 
 density distributions on a coarse grained level, but correlation effects 
 are completely missed. Simulation techniques give in principle exact 
 numerical results for the case of so-called primitive models where 
 the solvent is neglected and monomers and ions are modeled as
 charged soft spheres. Current open questions concern the structure of
 water in such dense systems: is water still described by a continuous
 medium with the bulk dielectric constant? Secondly, all approaches
 neglect the varying polarizability of the monomers and ions, which again
 give rise to pronounced image-charge effects in the brush geometry.
 Lastly, most brush systems have not reached their true equilibrium structure,
 so that we are in essence dealing with an non-equilibrium system. How 
 do we describe such non-equilibrium systems, and what governs the
 approach towards equilibrium including hydrodynamic and local friction 
 effects? We believe that the triad of scaling, field-theoretic, and simulation
 techniques will also allow to gain understanding of these more complicated
 issues in the near future.

\section{References}

\begin{figure}
 \epsfxsize=8cm
 \centerline{\vbox{\epsffile{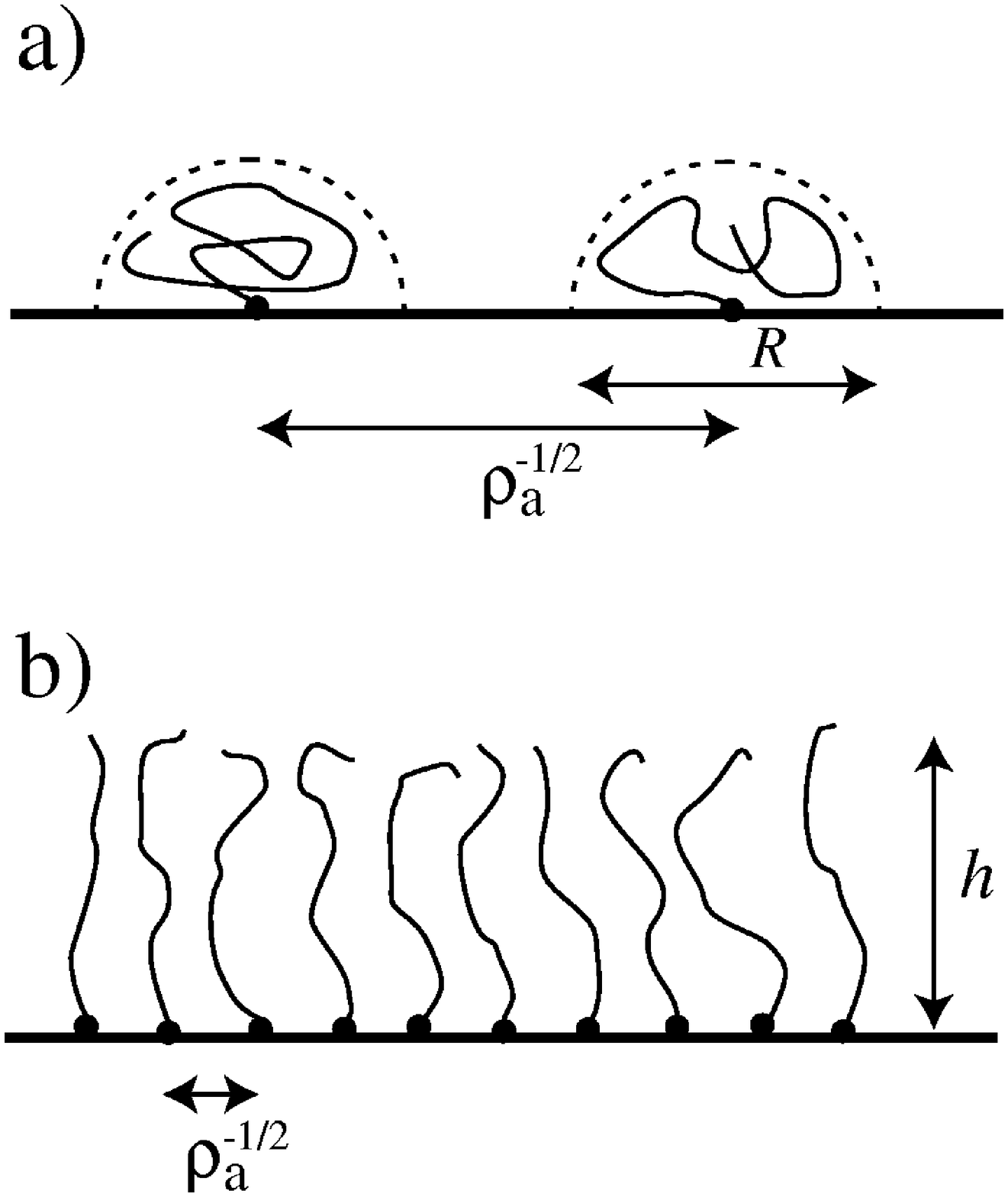} } }
\caption{ 
For grafted chains, one distinguishes between: a) the mushroom regime,
where the distance between chains, $\rho_a^{-1/2}$, is larger
than the size $R$ of a polymer coil; and, b) the brush regime, where
the distance between chains is smaller than the unperturbed coil
size. Here, the chains are stretched away from the surface due to
repulsive interactions between monomers. The brush height $h$
scales linearly with the polymerization index, $h \sim N$, and
thus is larger than the unperturbed coil radius $R$ which in the 
good-solvent regime scales according to Flory as $R_F  \sim a N^\nu$.}
\end{figure}

\begin{figure}
 \epsfxsize=8cm
 \centerline{\vbox{\epsffile{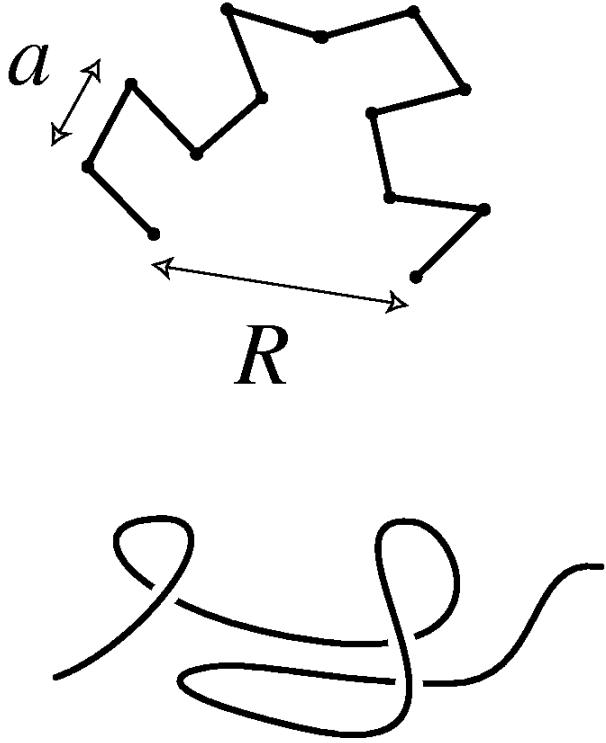} } }
\caption{ 
Top: Freely-jointed chain (FJC) model, where $N$ bonds of length $a$
are connected to form a flexible chain with a certain end-to-end distance $R$.
Bottom: In the simplified model, appropriate for more advanced theoretical
calculations, a continuous  line is governed by some
bending rigidity or line tension. This continuous model can be used when
the relevant length scales are much larger than the monomer size.
}
\end{figure}

\begin{figure}
 \epsfxsize=12cm
 \centerline{\vbox{\epsffile{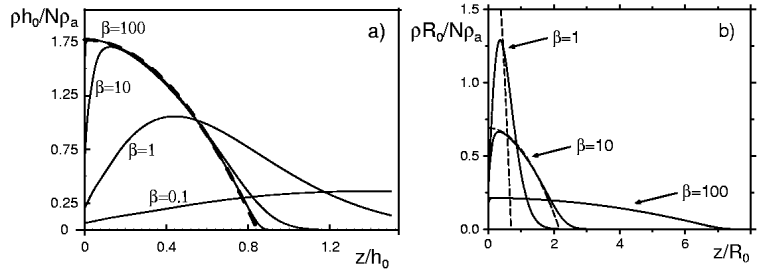} } }
\caption{  \protect
Self-consistent mean-field results for the density profile (normalized to unity)
of  a brush for different values of the interaction parameter $\beta$.
In a) the distance from the grafting surface is rescaled  by the 
scaling prediction for the brush height, $h_0$, and in b) it is
rescaled by the unperturbed polymer radius $R_0$.
 As $\beta$ increases, the density profiles approach the
parabolic profile (shown as dashed lines).}
\end{figure}

\begin{figure}
 \epsfxsize=12cm
 \centerline{\vbox{\epsffile{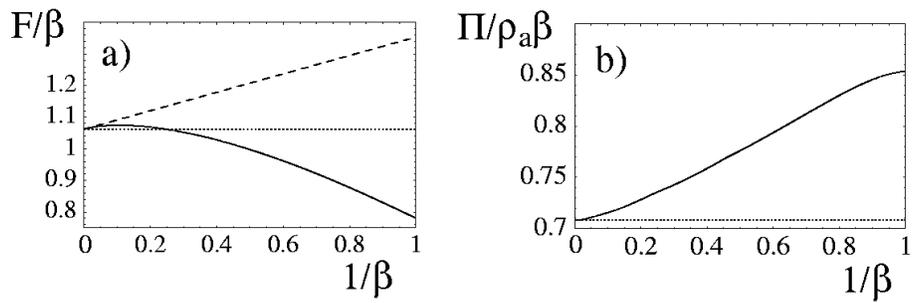} } }
\caption{  \protect
a) Mean-field result (solid line) for the rescaled brush free energy per polymer 
as a function of  the inverse interaction  parameter
$1/\beta$.  The infinite stretching result is indicated by a horizontal dotted line, 
the broken straight line denotes the infinite stretching result with the leading 
correction due to the finite end-point distribution entropy.
b) Rescaled lateral pressure within mean-field theory (solid line) compared
with the asymptotic infinite-stretching result (dotted line). 
}
\end{figure}

\begin{figure}
 \epsfxsize=10cm
 \centerline{\vbox{\epsffile{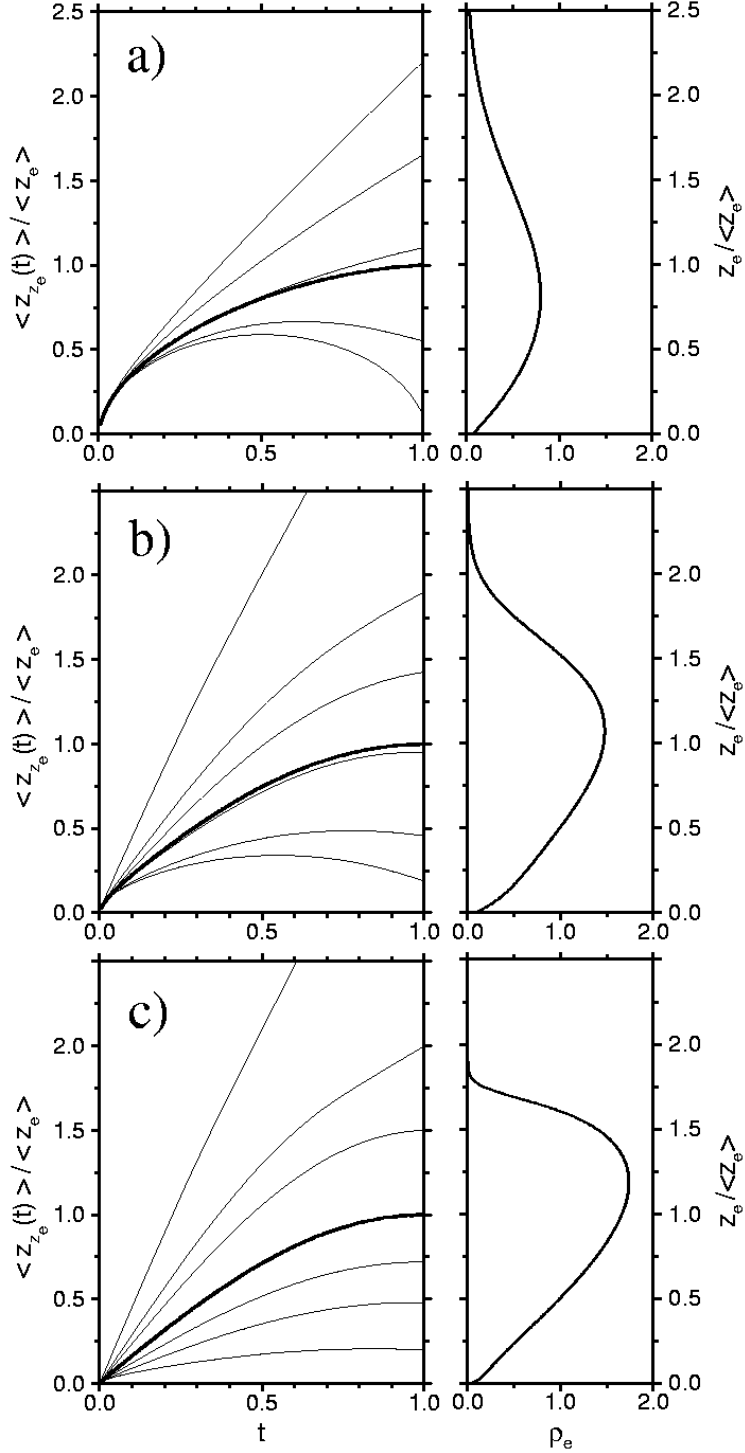} } }
\caption{  \protect
Left: Mean-field  results for the rescaled averaged polymer 
paths which end at a certain distance $z_e$ from the wall
for $\beta =\,\,$1, 10, 100 (from top to bottom), corresponding to 
stretching values of $\gamma=1.1, 1.9, 5.6$. The thick solid line shows 
the unconstrained mean path obtained by averaging over all end-point 
positions. Note that the end-point stretching is small but finite for all finite
stretching parameters $\beta$.
Right: End-point distributions.
}
\end{figure}

\begin{figure}
 \epsfxsize=12cm
 \centerline{\vbox{\epsffile{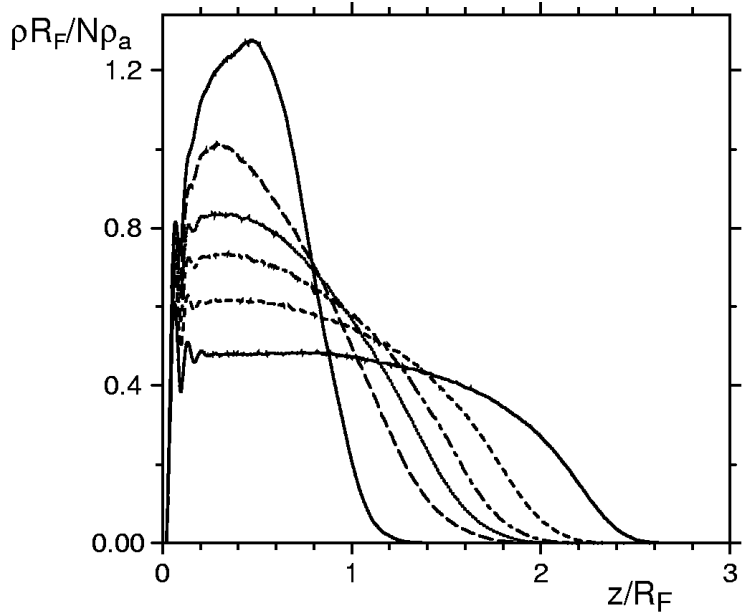} } }
\caption{  \protect Simulation results for the 
normalized monomer number density,
$\rho(z) R_F / N\rho_a$,  as a function of the scaled distance from
the grafting surface $z/R_F$ for anchored chains of length
$\,N=50\,$  and grafting densities $\,\rho_a \sigma^2$ = 0.02, 0.04,
  0.06, 0.09, and 0.17 (from top to bottom). Note that $R_F$ is determined
within the simulation for a single, free polymer chain.}
\end{figure}

\begin{figure}
 \epsfxsize=12cm
 \centerline{\vbox{\epsffile{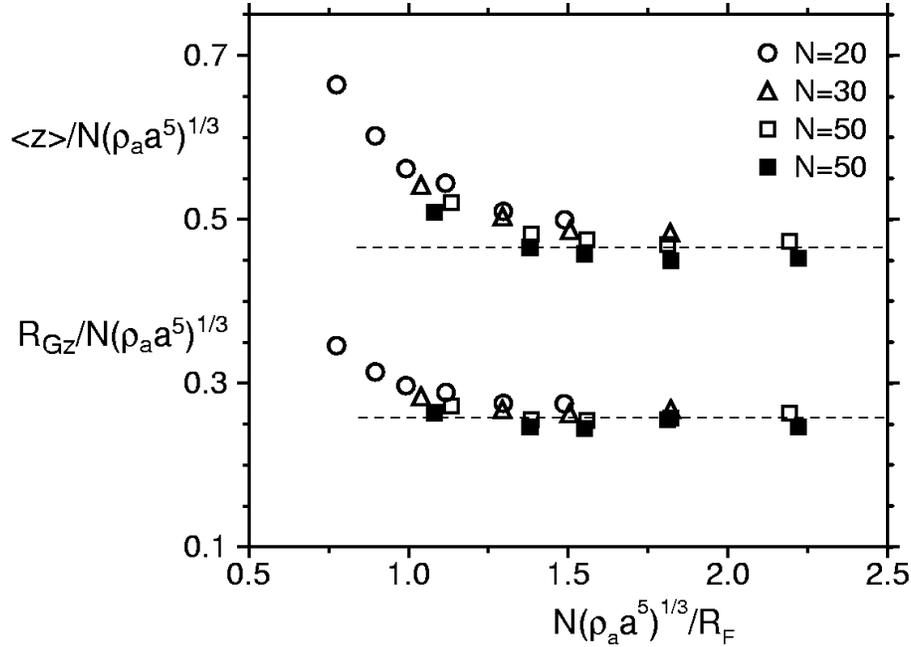} } }
\caption{  \protect Simulation results for the 
rescaled average monomer height
 $h/(N \rho_a^{1/3} a^{5/3})$ (top)   
 and the vertical component of the radius of gyration
  $R_{Gz}/(N\rho_a^{1/3}a^{5/3})$ (bottom) 
versus the scaling variable $N (\rho_a a^5 )^{1/3}/R_F$ for 
different chains lengths as indicated in the figure.
The open/filled symbols correspond to extensible and non-extensible
chains, respectively, which is controlled within the simulation by
the strength of the FENE bond potentials. As one can see,
for large values of the parameter combination $N (\rho_a a^5 )^{1/3}/R_F$
the data saturate at a plateau and are thus  in agreement with the scaling prediction.}
\end{figure}

\begin{figure}
 \epsfxsize=12cm
 \centerline{\vbox{\epsffile{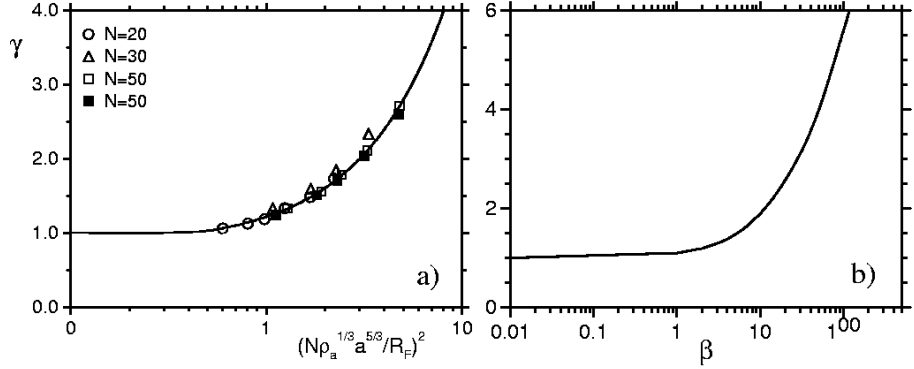} } }
\caption{  \protect
Effective stretching factor $\,\gamma = \langle z_e(\rho_a)/z_e(0) \rangle $ 
as obtained a) within the simulation 
as a function of the scaled anchoring density
    $(N\rho_a^{1/3} a^{5/3} / R_F)^2 $
and b) from the self-consistent field theory as a function
of the interaction  parameter $\beta$.
The comparison between the simulation results and the analytical results allows
to determine the effective $\beta$ parameter of a particular simulation set.
}
\end{figure}

\begin{figure}
 \epsfxsize=10cm
 \centerline{\vbox{\epsffile{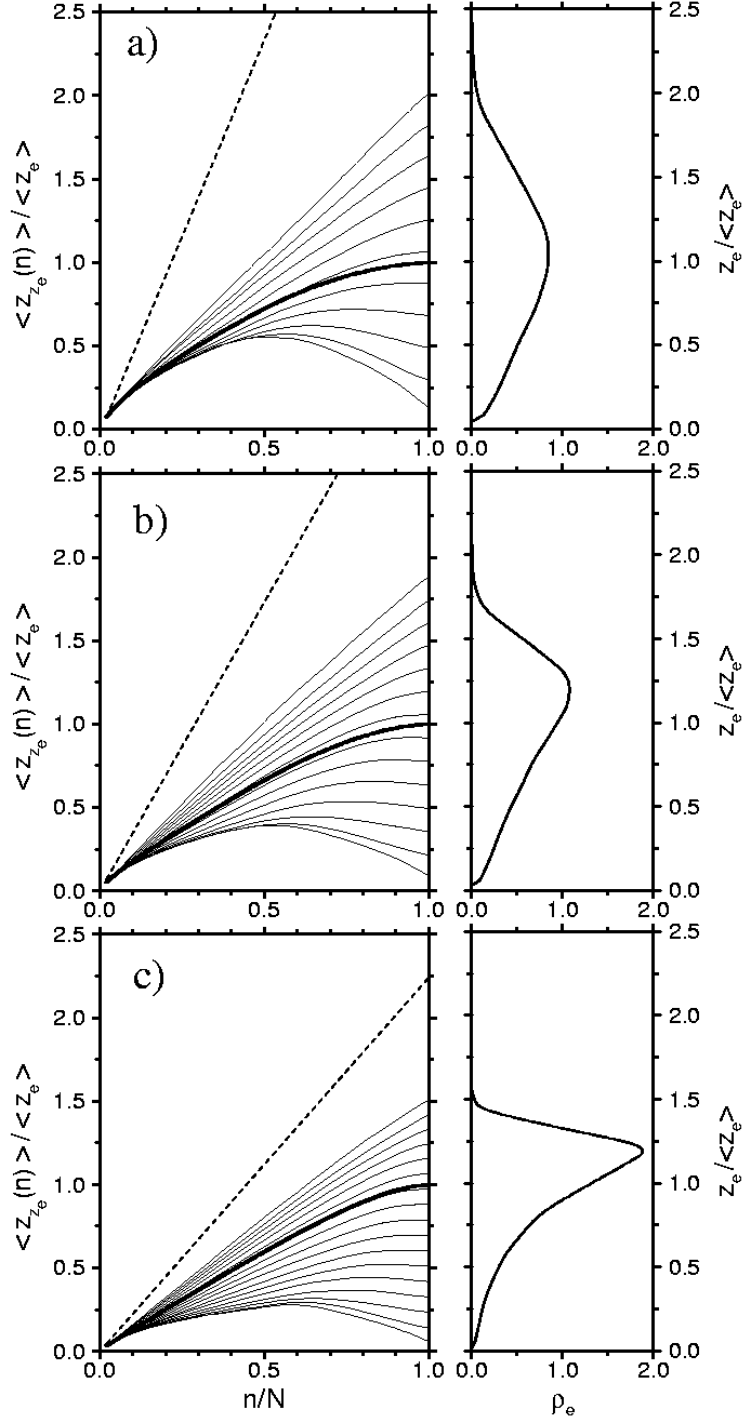} } }
\caption{  \protect
Left: Simulation  results for the rescaled averaged polymer 
paths which end at a certain distance $z_e$ from the wall
for $\rho_a \sigma^2  =\,\,$0.02, 0.06, 0.17 (from top to bottom), 
corresponding to 
stretching values of $\gamma=1.3, 1.8, 2.7$. The thick solid line shows 
the unconstrained mean path obtained by averaging over all end-point 
positions. Right: End-point distributions. All data are obtained for simulations
with 50 chains consisting of $N=50$ monomers each.}
\end{figure}

\begin{figure}
 \epsfxsize=8cm
 \centerline{\vbox{\epsffile{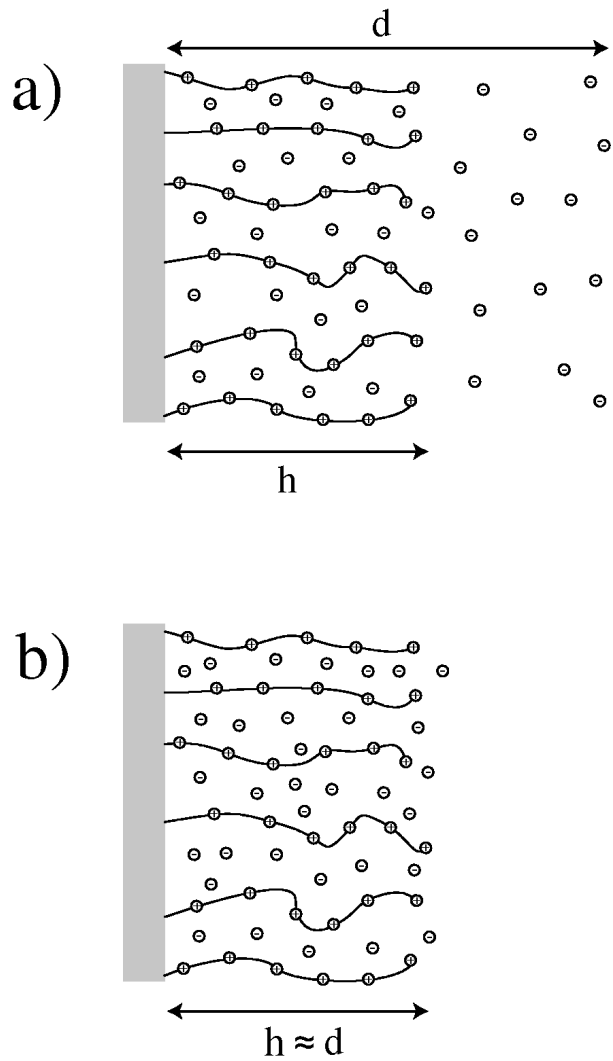} } }
\caption{ 
 Schematic PE brush structure. In a) we show the
weakly-charged limit where the counter-ion cloud has a thickness $d$
larger than the thickness of the brush layer, $h$. In  b) we show
the opposite case of the strongly-charged limit, where all
counter-ions are contained inside the brush and a single length
scale $d \approx h$  exists.}
\end{figure}

\begin{figure}
 \epsfxsize=10cm
 \centerline{\vbox{\epsffile{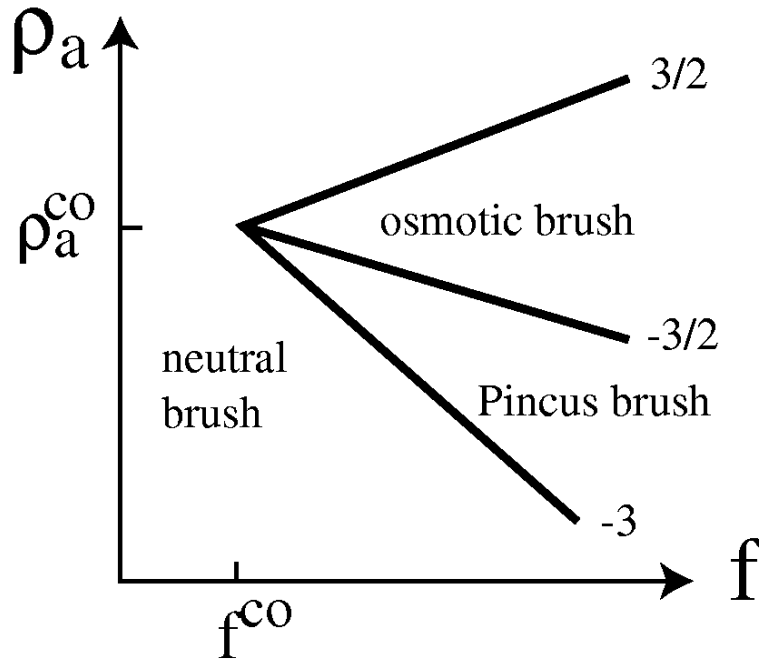} } }
\caption{ 
Scaling diagram for PE brushes on a
log-log plot as a function of the grafting density $\rho_a$ and the
fraction of charged monomers $f$. Featured are the Pincus-brush
regime, where the counter-ion layer thickness is much larger than
the brush thickness, the osmotic-brush regime, where all
counter-ions are inside the brush and the brush height is
determined by an equilibrium between the counter-ion osmotic
pressure and the PE stretching energy, and the neutral-brush
regime, where charge effects are not important and the brush
height results from a balance of PE stretching energy and
second-virial repulsion. The power law exponents of the various
lines are denoted by numbers.}
\end{figure}

\begin{figure}
 \epsfxsize=12cm
 \centerline{\vbox{\epsffile{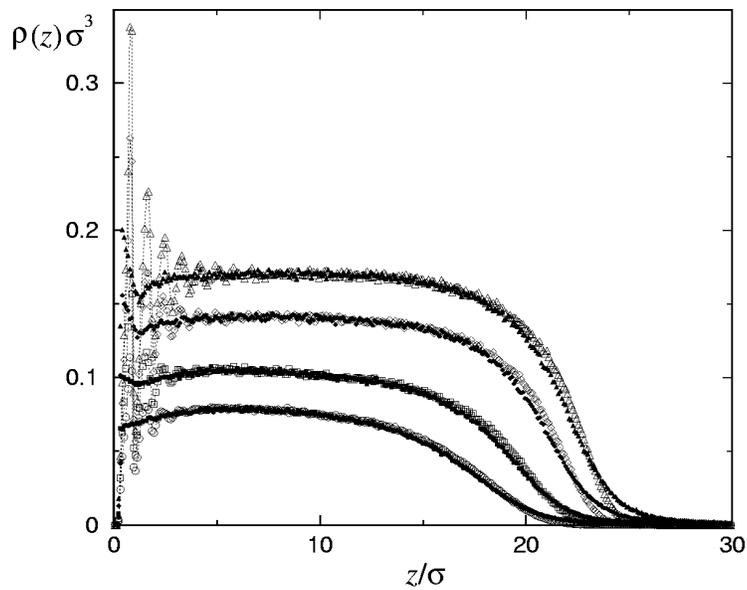} } }
\caption{  \protect
Density profiles of monomers  (open symbols) and
counterions  (filled symbols) as a function of the
distance from the anchoring surface. Shown are profiles for 
fully charged brushes of 36 chains of $N=30$ monomers at grafting densities (from bottom
to top) $\rho_{\mathrm{a}}\sigma^2 =0.042$
(circles),   $0.063$ (squares), $0.094$
(diamonds), and $0.12$ (triangles). As can be clearly seen, the counterion stay inside
the brush for all considered grafting densities and the local electroneutrality condition is 
satisfied very well.}
\end{figure}

\begin{figure}
 \epsfxsize=10cm
 \centerline{\vbox{\epsffile{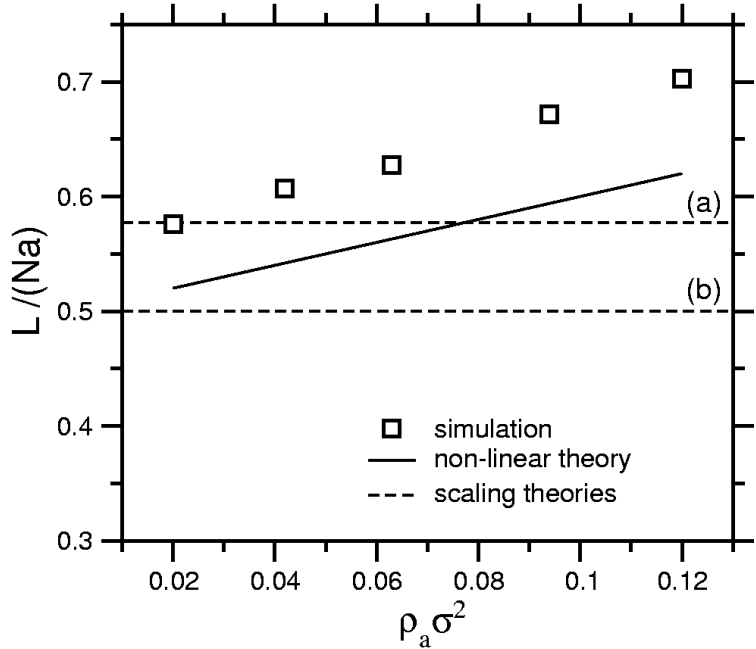} } }
\caption{  \protect The circles show simulation data for
the brush height as a function of grafting density for 
polyelectrolyte chains of $N=30$ monomers 
(contour length $L= 30 a$) with charge fraction of $f=1$.
The Bjerrum length is $\ell_B=\sigma$, which corresponds to an 
intermediate coupling strength quite relevant for fully charged chains in water,
and all ions are univalent.
The dotted lines (a) and (b) show the linear  scaling predictions, 
Eqs. (\ref{osmo}) and (\ref{osmo2}), 
with Gaussian and non-linear elasticity  respectively. 
The solid line shows the non-linear scaling prediction according to Eq. (\ref{osmo3}),
which includes non-linear elasticity as well as the finite excluded volume of
monomers and counterions.
}
\end{figure}

\end{document}